\documentclass[12pt]{iopart}

\usepackage{graphicx}
\usepackage{iopams}

\begin{document}

\title{Universal Magnetic Properties of 
sp$^3$-type Defects in Covalently Functionalized Graphene}

\author{Elton~J.~G.~Santos, Andr\'es Ayuela, Daniel S\'anchez-Portal} 
\address{Centro de F\'{\i}sica de Materiales (CFM-MPC),
Centro Mixto CSIC-UPV/EHU, 
Paseo Manuel de Lardizabal 5, 20018 San Sebasti\'an, Spain}
\address{Donostia International Physics Center (DIPC),
Paseo Manuel de Lardizabal 4, 20018 San Sebasti\'an, Spain}
\ead{\mailto{elton.jose@gmail.com},
\mailto{swxayfea@ehu.es}, \mailto{sqbsapod@ehu.es}}

\date{\today}

\begin{abstract}

Using density-functional calculations, we study the effect
of sp$^3$-type defects created by different covalent functionalizations 
on the electronic and magnetic
properties of graphene. 
We find that the induced magnetic
properties are {\it universal}, in the sense that they
are largely independent on the particular adsorbates
considered. 
When a weakly-polar single covalent bond 
is established with the layer, a local spin-moment 
of 1.0~$\mu_B$ always appears in graphene.
This effect is similar to that of H adsorption, 
which saturates one $p_z$ orbital in the carbon layer.
The magnetic couplings between the adsorbates show a
strong dependence 
on the graphene sublattice of chemisorption.
Molecules adsorbed
at the same sublattice couple
ferromagnetically, with an exchange interaction that decays 
very slowly with distance, while no magnetism is found for adsorbates 
at opposite sublattices. 
Similar magnetic properties are obtained if several $p_z$ orbitals
are saturated simultaneously by the adsorption of a large molecule.
These results might open new routes to engineer the
magnetic properties of graphene derivatives by chemical means.
 
\end{abstract}

\pacs{ 73.22.Pr, 73.20.Hb, 75.70.Ak, 75.75.-c}

\maketitle

\section{Introduction}

Graphene has exceptional electronic properties
with the potential to give rise to interesting 
new applications~\cite{Geim07,graph-review,Geim09_review}.
In particular, graphene is 
very attractive for spintronics~\cite{Wolf01}. 
This is mainly due to the very long spin relaxation and
decoherence times in these materials~\cite{Hueso07,Trauzettel07}.
There have been also numerous claims of the existence
of intrinsic magnetism in graphene associated with
the presence of 
structural defects~\cite{Esquinazi03,Ayuela03,Yazyev07,Palacios08,Guinea10,Yazyev10,Banhart11},
edge states~\cite{Son06,Yazyev08},
and partial hydrogenation of the layer.~\cite{Yazyev07,Lichtenstein08,Casolo09,Jena09} 
These observations suggest that graphene, besides 
being an ideal material
for spintronic circuitry, 
could also be used in active spintronic
devices. An interesting example is provided by graphene nanoribbons.
The energies of the polarized states 
localized at the
edges of zigzag nanoribbons or nanotubes
can be controlled by external
electric fields, and this effect can be used to design
a spin-filtering device.~\cite{Son06,Ayuela08}
However, in spite of the growing interest, to date there 
has not been a conclusive experimental 
confirmation of this intrisic magnetism in graphenic materials.
While some experimental groups claim to have found clear
indications of ferromagnetism
in irradiated graphite~\cite{Esquinazi03}, graphene prepared
from reduced graphene oxide~\cite{Wang09}
and hydrogenated graphene~\cite{Xie11}, these results are 
challenged by other groups with alternative 
experimental evidences~\cite{Sepioni10,Asenjo10}. 

One of the main difficulties for the observation of ferromagnetism
in defective and hydrogenated samples might be related
to the bipartite lattice of graphene.
It is theoretically well known, at least within a simplified description 
based on a $\pi$-tight-binding Hubbard model, 
that defects created in different sublattices tend to couple 
antiferromagnetically~\cite{Lieb,Palacios08,Roche11,Roche11_bis}. 
Since, for example, defects created by irradiation are expected to 
be randomly distributed, this would prevent the formation of 
ferromagnetic order. Therefore, 
to generate ferromagnetism in graphene and other graphenic 
nanostructures, 
it is desirable to create {\it all} the defects or to adsorb {\it all} 
the hydrogen atoms in
{\it one} of the graphene sublattices.~\cite{Jena09}
Unfortunately, technically this might be very difficult to achieve.
For this reason, it is interesting to look for alternative
ways to induce magnetism in the carbon layer. In the present work
we analyze, using first-principles electronic structure calculations,
the role of different 
sp$^3$-type defects created by covalent functionalization
as a source of magnetism in graphene.
Covalent functionalization
has not been frequently investigated 
in this context, even though it has been 
used for a long time
in the chemistry of carbon nanostructures and 
now in graphene~\cite{Hirsch02,Ruoff09,Loh10}.
We find that, under very 
general conditions, any molecule attached to the carbon layer
through a weakly-polar single bond produces an effect similar to that of 
hydrogen adsorption, i.e., induces a significant spin moment in
the system. In particular, we explicitly show that the 
induced magnetism 
is largely independent on the size of the
adsorbed molecule and even on its chemical or biological activity. 
Regarding the coupling of the induced moments, we
show that identical results
are obtained
by the simultaneous adsorption of several molecules
or by the adsorption of a larger molecule that creates
several bonds with the layer.
This simple fact may open new and interesting routes to tune the
magnetism of graphene. In principle,
molecules can be produced with the appropriate
structural and chemical characteristics to create
bonds with the layer according to predefined
patterns. Thus,
the powerful 
techniques of organic and surface chemistry could  
be applied to synthesize magnetic derivatives of
graphene that behave according 
to well studied theoretical models~\cite{Lieb,Palacios08,Roche11,Roche11_bis}. 
This is particularly attractive due to the recent 
sucessful synthesys
of different graphene derivatives using surface chemical 
routes~\cite{Mullen10,Mullen10_bis}.
Thus, the synthesis of carbon nanostructures with functional groups
at predefined positions, starting from previously functionalized
monomers, seems plausible nowadays. Our study deals with the
magnetic properties of such nanostructures. 

In more detail, 
our calculations predict that, when a single C$-$C covalent bond is 
established between an adsorbate and graphene,   
a spin moment of 1.0 $\mu_B$ 
is always induced in the system. The size and spatial distributions of
these moment are nearly 
independent of the particular adsorbate. 
We explicitly show that this effect occurs for a 
wide class of organic and inorganic molecules with different 
biological and chemical activity, 
e.g. alkanes\cite{Zettl08}, 
polymers\cite{Ramanathan08,Nutt09,Liu08}, 
diazonium salts\cite{Tour10,Tour09,Tour08}, 
aryl and alkyl radicals\cite{deHeer10,Cao10}, 
nucleobases\cite{Rao09,Bianco09}, 
amide and amine groups\cite{amide10}, glucose,\cite{Saha10}
and organic acids\cite{Tour08,Ruoff07}. While a similar behavior
is observed for hydrogen adsorption, we show for other
adsorbates 
(NH$_2$~\cite{amino11}, 
OH~\cite{Ruoff07} and F~\cite{Fluorographene10,Zhu10}) 
that the induced spin-moment decreases with increasing 
bond-polarity. Moments induced by adsorption 
in the same 
sublattice align ferromagnetically, with an exchange coupling
that falls off very slowly with the distance between adsorbates 
($\sim r^{-(1+\epsilon)}, \epsilon\sim 0.20$).
In contrast, for molecules adsorbed on opposite 
sublattices no magnetic solutions 
could be stabilized. 

\section{Methods}

Our first-principles electronic structure calculations use
density functional theory~\cite{kohn1965} as
implemented in the SIESTA code~\cite{soler02}.
We use the generalized gradient approximation\cite{pbe-functional},
norm-conserving pseudopotentials\cite{troullier91} and
a basis set of numerical atomic orbitals~\cite{soler02}.
A double-$\zeta$ polarized (DZP) basis set has been used
for the calculation of the magnetic moments and
electronic band structures of all our systems.
We have checked that the relaxed structures with a DZP basis
are almost identical to those obtained using a double-$\zeta$ (DZ) basis.
Therefore, we have used the smaller DZ basis for the relaxations
of systems containing more than 100 atoms.
The force tolerance for structural optimizations was 0.04~eV/\AA.
The integration over the Brillouin zone was performed using
a well converged $k$-sampling
equivalent to $136\times136\times1$ k-points
for the unit cell of graphene.
The fineness of the real-space grid used
by SIESTA was equivalent to a 150 Ry plane-wave cutoff.
We have repeated some calculations with the VASP
code\cite{kresse93, kresse96} using a well-converged plane-wave cutoff
energy of 400 eV combined with the projected-augmented-wave (PAW) method.
Other computational details were as previously described.
The use of PAW potentials allows to check the inherent limitations
of the norm-conserving pseudopotentials. The
two methods provide almost identical results.
We have checked that, when a sufficiently dense k-sampling is used,
our results are very weakly dependent on
the supercell size. In particular, a spin moment of 1~$\mu_B$ is
always associated with the formation of single
C-C bonds with graphene, as well as with H adsorption.
All the
calculations presented in the text
used a 8$\times$8 supercell of graphene.

\section{Results and discussion}

\begin{figure}
\begin{center}
\includegraphics[width=3.500in]{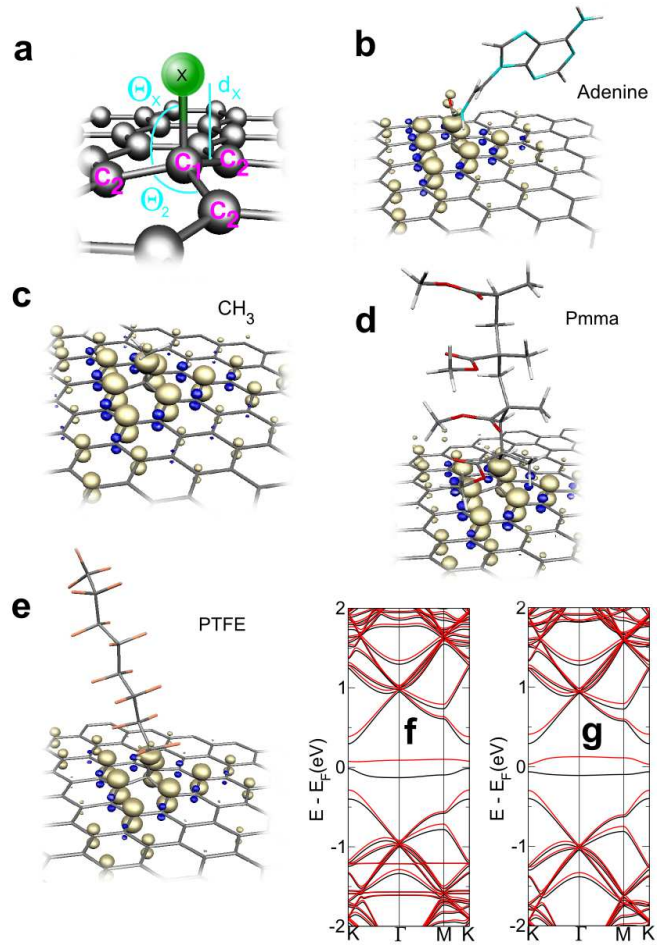}
\end{center}
\caption{(Color online) (a) Schematic representation 
of the adsorption geometry of our systems and
the nomenclature used in Table~\ref{table1}.
X denotes here the atom directly bonded to graphene.
Panels (b)-(e) show the isosurfaces ($\pm$0.019 e$^-$/Bohr$^3$)
of the magnetization density 
induced on graphene by functionalization using 
adenine derivatives\cite{Bianco09}, 
CH$_3$, PMMA and PTFE.
Majority and minority spin  
densities correspond respectively to light and dark surfaces, which 
localize in different graphene sublattices and show
a slow decay with distance in all the cases.
Panels (f) and (g) show the spin polarized band 
structures of a 8$\times$8 graphene supercell with a single 
adenine derivative
and a CH$_3$ molecule chemisorbed on top of one carbon atom, 
respectively. 
The black and red (gray) lines denote 
the majority and minority spin bands, respectively.
The Fermi energy is set to zero.}
\label{fig1}
\end{figure}

We first calculated the adsorption geometry of several molecules
that can establish 
single bonds with the graphene layer. 
Most of them are attached to graphene through
a single C$-$C covalent bond (denoted by X$=$C in Table~\ref{table1},
X being the atom that bonds to graphene).
We have also considered 
other adsorbates (H, NH$_{2}$, OH and F) 
that form heteropolar bonds with graphene.
In all cases the adsorbate sits directly on top
of one of the C atoms of the layer [see
the scheme in Fig.~\ref{fig1}~(a)]. This central C atom (C$_{1}$) moves
vertically and approximately keeps a
symmetric three-fold structure respect to its neighbors (C$_2$). 
Table~\ref{table1} describes
adsorption geometries of the different molecules. 
The angles $\Theta_{X}$ and $\Theta_{2}$ are in the range 
$102.1^{o}-107.0^{o}$ and $111.8^{o} - 115.7^{o}$, respectively.
These values are half-way between those corresponding to 
a $sp^2$ hybridization of the central C$_1$ atom
($\Theta_{X} = 90^{o}$ and $\Theta_{2} = 120^{o}$), and those 
expected for a $sp^3$ local hybridization 
($\Theta_{X} =\Theta_{2} = 109.5^{o}$). 
Thus, the observed geometries can be understood
by a slight modification of the sp$^2$
hybridization of C$_1$ that gains 
certain $sp^3$ character~\cite{Lichtenstein09}. 
The largest $sp^3$ character among the studied molecules
corresponds to the adsorption of nitrobenzene (C$_6$H$_4$NO$_2$), 
anisole (C$_6$H$_4$OCH$_3$), PMMA, and polystyrene groups. 
The smallest local distortion 
of the carbon layer corresponds
to F adsorption, reflecting a larger ionic character of the interaction.

The main results for the magnetism of all the studied adsorbates are 
presented in Fig.~\ref{fig1} and Table~\ref{table1}. 
We first focus on the adsorbates attached to the layer by 
a single C$-$C bond. 
In this case, the graphene-adsorbate complexes always
exhibit a spin moment of 1~$\mu_B$. 
The induced spin polarization texture is shown
in Fig.~\ref{fig1} for the adsorption of the groups adenine (b) and
methyl (c), and the PMMA (d)
and PTFE (e) polymers. The calculated patterns
are remarkably similar in all the cases. Surprisingly, although the
spin moment is induced by the functionalization,
it is mostly localized in the graphene layer. The $p_z$ character
of the electronic states behind the spin polarization is also
evident in the plots.
The distribution of the spin moment follows the
bipartite character  of the
graphene lattice: C atoms
in the opposite (same) sublattice than the saturated C atom
C$_1$ show majority (minority) spin polarization.
Although the total spin moment is 1.0~$\mu_{B}$, 
a Mulliken analysis only assigns 
$\sim$0.10~$\mu_{B}$ to the C atoms that form
the graphene-adsorbate bond (C$_{1}-$X, with X=C).
In contrast, the contribution from the three first
nearest-neighbors [C$_2$ in Fig.~\ref{fig1}~(a) ]
is 0.34~$\mu_{B}$,
-0.13~$\mu_{B}$ from the next nearest-neighbors, 
0.26~$\mu_{B}$ from the third neighbors,
and 0.40~$\mu_{B}$ integrated over larger distances. This clearly 
shows the slow decay of the spin moment
induced in graphene by this type of covalent functionalization. 
The latter is also reflected in the long-range exchange interactions
between adsorbates that we describe below. 

\begin{figure}
\begin{center}
\includegraphics[width=3.600in]{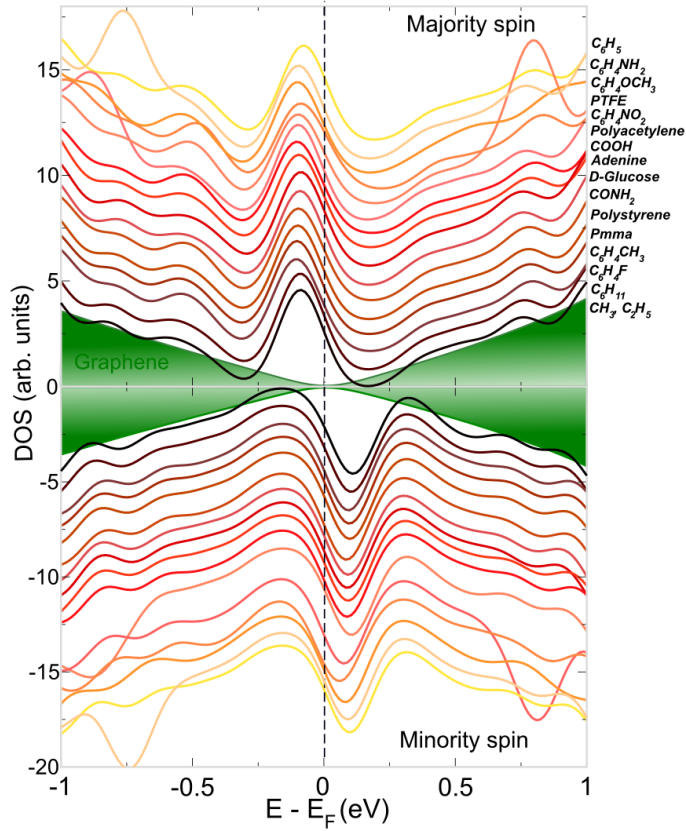}
\end{center}
\caption{(Color online) Spin-polarized density of states 
for one molecule chemisorbed on a graphene supercell. All the
adsorbates considered
attach to the layer through a single C$-$C bond.
For clarity, curves for different 
adsorbates have been shifted and smoothed with 
a Lorentzian broadening of 0.12 eV.
The Fermi level is marked by the dashed line and is set to zero.
The shaded regions denote the density of states of pristine graphene.
}
\label{fig2}
\end{figure}

In order to understand the origin of this spin polarization we 
have explored
the band structure of the functionalized layers.
Figure~\ref{fig1}(f) and (g) show the band structure  
for graphene functionalized with 
adenine and methyl groups, respectively, as representative cases.
The magnetization comes from a very narrow 
defect state that is pinned at the 
Fermi level (E$_{F}$). This state shows a 
predominant $p_z$  contribution from the nearest neighbors 
C$_{2}$ to the
defect site, with a much smaller component from the adsorbate itself
and a negligible contribution from the central carbon atom C$_1$.
This state is reminiscent of the defect level that appears
associated with a carbon vacancy  
in a $\pi$-tight-binding description of graphene~\cite{Palacios08}.
The origin of the level is also similar in the present case, 
since the new C$-$C 
bond formed upon
adsorption saturates the $p_z$ orbital at C$_1$.
The small  dispersion  of  this  $p_z$-defect band combined  with  its  
partially filled character favors the spin-uncompensated solution.
Table~\ref{table1} shows the spin-splitting of the 
$p_z$ defect-band ($\delta$E$_{s}$) at $\Gamma$ 
point~\cite{note1}.
This splitting varies in the small range 0.19-0.24~eV for all the adsorbates 
anchored by a single C$-$C bond (see also Fig.~\ref{fig2}). 
This further confirms the similar character 
and localization of the defect state
behind the spin polarization for all the molecules.
The energy gain with 
respect to the spin-compensated solution 
($\Delta$E$_{M}$ = E$_{PM}$-E$_{FM}$) is comparable
for all adsorbates, being
$\sim$45~meV.

The observed spin moment is localized in graphene and, 
therefore, it is derived from the electronic
states of the carbon layer. However, the spin moment appears
due to the 
covalent functionalization with
non-magnetic molecules. As mentioned above, the adsorbate
saturates the p$_z$-orbital of one of the C atoms in
the layer and, thus, creates a defect analogous to a $\pi$-vacancy.
Still, according to the results in Table~\ref{table1}, 
the polarity of the graphene-adsorbate
bond is an additional ingredient 
that determines the size of the spin 
moment of the system~\cite{note2}. 
As the ionic character of the bond increases 
(going from H to F in Table~\ref{table1}) 
we observe a transition from a magnetic 
to a non-magnetic ground state. We can use the results from Mulliken 
population analyses to characterize the relative polarity of these bonds. 
Although the absolute values of the charge transfer 
have to be taken with care, since they are known to 
be strongly dependent on the size and 
quality of the basis set, results obtained with the same type of basis 
set can be compared and used to establish trends~\cite{Daniel}. 
In our calculations both H and NH$_2$ radicals present 
almost negligible charge transfers
of 0.003 and 0.046 electrons from graphene, respectively. 
Consequently, H and NH$_2$
behave similarly to the adsorbates presented previously, 
i.e., those bonded to graphene through homopolar C$-$C bonds.
The electronic band-structures induced by H and NH$_2$ are 
similar to those presented in Fig.~\ref{fig1}, and
solutions with large spin-moments close to 1~$\mu_B$ 
are also stabilized. 
In contrast, 
the charge transfers from graphene to OH and F increase to
0.26 and 0.27 electrons, respectively. 
As a consequence, 
the Fermi level and the p$_z$ defect-band pinned to it appear now
0.2-0.3~eV below the Dirac point.
This is consistent with
the appreciable charge transfer towards the 
adsorbate and the consequent doping of the layer with holes.
The additional population of the defect level, 
as well as its energy position in a region 
with a larger density of states of graphene, causes 
a drastic reduction of the spin moment for OH and a non-magnetic ground state 
for F. 
However,  
calculations using the fixed spin-moment method, presented in 
parentheses in Table~\ref{table1}, show that the energy
penalty needed to develop a spin solution is quite small even
in the case of F ($\Delta$E$_M$=$-$24.1~meV).
Thus, although non-magnetic according to our calculation, 
the graphene$-$F bond
can be easily polarized (at least in the low adsorbate-density
regime explored here). 
This can be important to interpret the recent
experimental report of colossal magnetoresistance 
in this system.~\cite{Hong11} 

\begin{table}
\caption{Results of the structural parameters for all the
studied adsorbates. We follow Fig.~\ref{fig1}~(a)
for the nomenclature.
Energy gain $\Delta$E$_{M}$ due to the spin polarization, total spin moment S 
and spin-splitting $\delta$E$_{s}$ of the defect level are also included.
The numbers in parentheses for F are calculated using the fixed-spin method.
}
\begin{center}
\begin{tabular}{lcccccc} 
\hline
& d$_{X}$(\AA)&$\Theta_{X}$($^o$)&$\Theta_{2}$($^o$)& 
$\Delta$E$_{M}$(meV) &S($\mu_{B}$) & $\delta$E$_{s}$(eV)\\ \hline 
X = C   & & & & & &        \\
CH$_3$          & 1.59  & 104.5 & 113.9  & 48.6    & 1.00  & 0.23 \\
C$_2$H$_5$      & 1.59  & 105.0 & 114.0  & 48.0    & 1.00  & 0.23 \\
C$_6$H$_{11}$   & 1.67  & 105.9 & 112.5  & 47.4    & 1.00  & 0.23  \\
C$_{6}$H$_{5}$  & 1.59  & 105.7 & 112.9  & 46.2    & 1.00  & 0.23 \\
C$_{6}$H$_{4}$F & 1.61  & 106.6 & 112.2  & 45.6    & 1.00  & 0.23  \\
C$_{6}$H$_{4}$NO$_2$&1.60&107.0 & 111.8  & 32.3    & 1.00  & 0.20  \\
C$_{6}$H$_{4}$OCH$_3$&1.59&106.9& 111.9  & 38.4    & 1.00  & 0.21  \\  
C$_{6}$H$_{4}$CH$_{3}$&1.61&106.6&112.2  & 46.6    & 1.00  & 0.23  \\
C$_{6}$H$_{4}$NH$_{2}$&1.60&106.5&112.3  & 42.8    & 1.00  & 0.22 \\
CONH$_{2}$     & 1.65& 104.9 & 113.6     & 43.2    & 1.00  & 0.21  \\
COOH           & 1.60& 104.5 & 113.6     & 41.6    & 1.00  & 0.22  \\
PMMA           & 1.67& 106.8 & 112.0     & 47.1   & 1.00  & 0.23  \\
PTFE           & 1.67& 105.8 & 112.9     & 64.8   & 1.00  & 0.19  \\
Adenine derivative$^*$ & 1.65&104.7  & 113.8     & 42.5   & 1.00  & 0.23  \\
D-Glucose      & 1.64& 105.1 & 113.5     & 41.4   & 1.00  & 0.21  \\  
Polystyrene    & 1.60&106.8  & 112.0     & 40.0    & 1.00  & 0.21  \\
Polyacetylene  & 1.58& 105.0 & 113.6     & 50.1   & 1.00  & 0.24       \\ \hline
X = H, N, O, F    &     &        &            &         &       &       \\ 
H              & 1.12& 102.6 & 115.3     & 46.5   & 1.00  & 0.24  \\
NH$_{2}$       & 1.52& 105.2 & 113.4     & 27.7  & 0.89  & 0.20 \\
OH             & 1.52&  103.7& 114.6     & 8.4   & 0.56  & 0.12  \\
F              & 1.55& 102.1 & 115.7     &  0.0   & 0.00   & 0.00   \\
               &     &        &            &(-24.1)&(1.00) &(0.16)  \\ \hline
\end{tabular}

$^*$ 9-(2-aminoethyl)adenine anchored to a carbonyl group~\cite{Bianco09}

\label{table1}
\end{center}
\end{table}

We now look at the electronic structure around E$_F$ in more detail. 
Fig.~\ref{fig2} shows the calculated density of states (DOS) per spin channel 
for different adsorbates chemisorbed on graphene through a C$-$C bond.
Despite the curves being shifted and 
smoothed with a Lorentzian broadening, the data 
collapse onto a single pattern: one fully polarized peak appears close 
to E$_F$. This confirms the universality of the spin moment 
associated with this type of covalent functionalization, 
independent of the particular adsorbate.
In particular, our results point out to an almost perfect analogy
between chemisorbed hydrogen~\cite{Yazyev07,Lichtenstein08,Palacios08}
and adsorbates bind to graphene through single C$-$C bonds, which
is not an obvious behaviour.
Deviations (not shown in Fig.~\ref{fig2})
from this collapsed curve 
are observed for adsorbates that form with graphene 
bonds of a considerable different character and polarity.

\begin{figure}
\begin{center}
\includegraphics[width=5.000in]{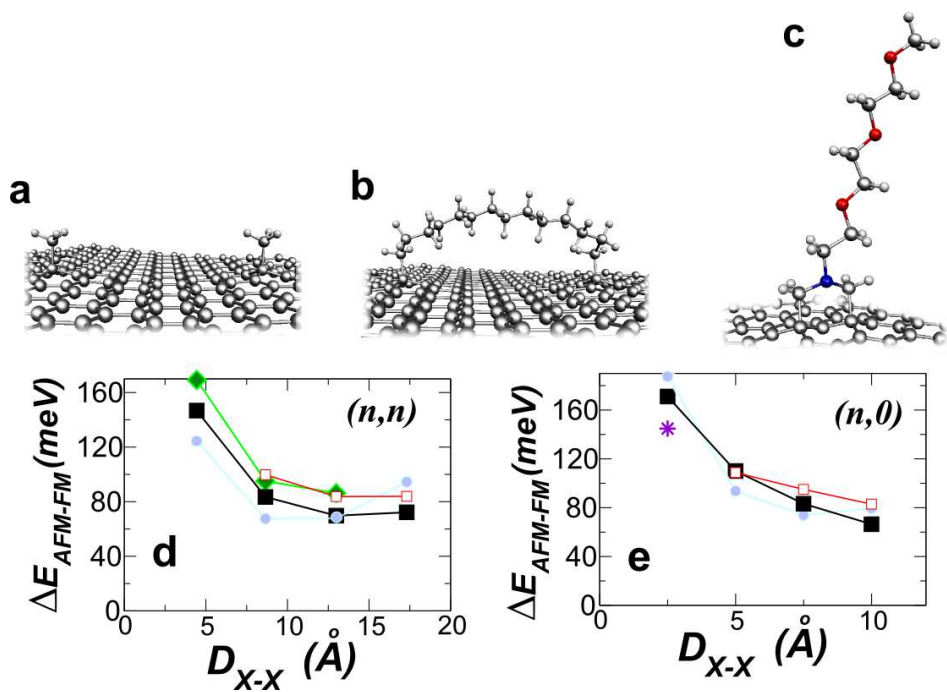}
\end{center}
\caption{(Color online) 
Schematic 
representation of some of the systems used 
to study the exchange coupling between adsorbate-induced 
spin moments in graphene: (a) two methyl groups,
(b) the two ends of
an alkane chain attached
at two neighboring locations, and (c) the two bonds formed in a
cycloaddition.
The calculated energy difference, using a 8$\times$8 supercell, between
antiferromagnetic and ferromagnetic alignments of the local
spin-moments are shown as
a function of the distance along the armchair (d) and zigzag (e)
directions. 
Only data for saturation at the 
the same graphene sublattice (AA configurations) are shown. 
Simultaneous 
adsorption at different sublattices (AB configurations) always gave rise 
to non-spin-polarized solutions. 
Filled and empty squares correspond, respectively, to H and CH$_3$.
Diamonds (green symbols) in panel (d) stand for the alkane chain and 
the star in panel (e) for
the hypothetical AA-cycloaddition shown in (c). 
Circles (gray symbols) correspond to the best fit using a Heisenberg model (see text).}
\label{fig3}
\end{figure}

Next we study the magnetic couplings between adsorbates
at low concentrations. Figure~\ref{fig3} shows the
energy differences between ferromagnetic (FM) and antiferromagnetic (AFM)
alignments of the adsorbate-induced spin moments as a function
of the distance between the adsorption sites. These results can 
be used to estimate the size and the 
distance dependence of the exchange couplings
between the adsorbate-induced moments in graphene. We have studied
systems of two different kinds: in one case two
independent molecules are adsorbed at different locations, while in the
other case
a single molecule forms two covalent bonds with  
the graphene layer. To exemplify the first type of system, we have used
H and CH$_3$ as representative adsorbates.
Several observations can be made in this case: ({\it i}) If the two molecules 
are located  at the  same sublattice (AA adsorption), 
the FM alignment is always
more stable than the AFM one. In the FM case, the total spin
moment integrates to  2.00 $\mu_{B}$ for both H and CH$_{3}$ and
the local spin  population remains  
nearly constant around each defect site, roughly insensitive
to the presence of the neighboring adsorbate;
({\it ii}) If the two molecules are at the different 
sublattices (AB adsorption),
we could not stabilize any magnetic solution, and the system
converges to a spin-compensated solution with no local moment;
({\it iii}) The size and distance dependence of the magnetic
couplings are very similar for H and CH$_3$, which further supports
the analogy presented above;
({\it iv}) For two methyl groups at short distances ($<$ 5.0~\AA) we
could only find FM solutions.

We have also considered the magnetic couplings between local
spin-moments induced in the layer by the formation of two
single C$-$C bonds between graphene and the {\it same} molecule. 
We first explore this situation using a long alkane chain attached to 
graphene at two different locations [see Fig.~\ref{fig3}~(b)].
Since the characteristics of the
spin moment induced by 
covalent functionalization are quite 
independent of the adsorbed molecules, as far as they are 
anchored by a single C-C bond, 
we expect in this
case a result similar to that previously found for two 
methyl groups.
This is confirmed by the results in Fig.~\ref{fig3}~(d). For AA
adsorption, i.e., when the two bonds between the 
alkane chain and graphene are established with atoms in the same
graphene sublattice, the 
FM alignment with a 2~$\mu_B$ total spin moment 
is still the most stable.
The size of the exchange couplings is almost identical to the case
of two independent methyl groups. The only significant
difference is that for the alkane chain we succeeded to find
an AFM configuration at short distances. Furthermore, in the present case
AB adsorption also converges to spin-compensated solutions.

Figure~\ref{fig3}~(c) and (e) show another example of magnetism
associated with several
bonds formed by the same molecule. In this case, we have
considered a dipolar cycloaddition~\cite{Prato10,Houk11}.
Usually this kind of cycloaddition saturates simultaneously
two first-neighbor carbon atoms in graphene, i.e., it corresponds
to an AB configuration. As a consequence, in agreement with
the results presented so far, the cycloaddition does not
create a spin moment in the layer. However, we have
considered here another configuration in which the
molecule is attached to two second-neighbor carbon atoms in 
graphene. This artificial AA-cycloaddition is 1.18~eV less stable
than the standard AB-cycloaddition. In 
accordance to the universality of our results so far, 
the AA-cycloaddition gives rise to a spin-polarized solution. 
As expected, the FM solution with 2$\mu_B$ spin moment is more stable, and
the estimated exchange coupling fits well with those found
for H, the methyl radical and the alkane chains.

The observed magnetic behavior follows closely the
expectations based on Lieb's theorem for a bipartite
Hubbard model at half-filling~\cite{Lieb}, which is an appropriate
model to describe the low energy electronic excitations
in graphene. Lieb's theorem predicts that graphene
should develop a {\it total} spin moment of 2$\mu_B$ for
AA adsorption and zero for AB adsorption~\cite{Lieb,Palacios08}.
However, this does not explain why a spin-compensated solution
is more stable than an AFM solution for AB adsorption, as we
find in our first-principles calculations.
The reason for this behavior can be rationalized as
follows.
When adsorption takes place at opposite sublattices, due to
the bipartite character of graphene, 
the interaction between defect levels in neighboring
adsorption sites is appreciable. 
This interaction opens a bonding-antibonding
gap in the $p_z$-defect bands that contributes to 
stabilize the spin-compensated solution. 
The non-magnetic solution is favored when 
this gap is 
larger than the spin splitting $\delta$E$_{s}$ of the
isolated defect~\cite{Co-paper10}. 
A detailed analysis of the calculated band structures 
confirms this interpretation.
It is noteworthy that a similar 
behavior was also observed for 
chemisorbed hydrogen~\cite{Yazyev07,Lichtenstein08}, 
vacancies~\cite{Kumazaki07,Palacios08}, 
and even for substitutional Co atoms 
in graphene~\cite{Co-paper10}. All of them 
exhibit a very similar electronic structure nearby E$_F$.
According to our interpretation, at
sufficiently large distances
between the adsorbates,
a magnetic configuration with non-zero local spin-moments 
at each adsorption site should become favorable even for
AB adsorption. 
At such long distances,
the gap in the defect band becomes smaller
than $\delta$E$_{s}$ and the local spin-moments are recovered. However,
in accordance with  Lieb's theorem,
an AFM configuration will be favored in this case.
The necessary 
distances to stabilize these AFM solutions are larger than those
explicitly considered in our {\it ab initio} calculations presented
in Fig.~\ref{fig3}. 
The co-existence of non-magnetic
and AFM solutions for AB configurations has been observed
using first-principles calculations
in the case of substitutional Co atoms in graphene~\cite{Co-paper10}.

An important consequence of the
results presented above is that
it is possible to turn graphene magnetic 
by chemical functionalization~\cite{Xie11}.
Unfortunately, they also tell us that, in order to achieve and control 
such magnetism,
it is necessary to have selectivity on the adsorption site, i.e.,
molecules should preferentially attach to one of the graphene sublattices.
This can be difficult to achieve in practice, particularly 
if we take into account that AA configurations are usually less
favorable than AB configurations~\cite{Lichtenstein08,Casolo09}. 
Therefore, some difference must be
introduced between A and B sites to promote adsorption on 
a single sublattice. For example, one possibility 
could be the
registry with an appropriate substrate, but this
can also alter other desired properties of the graphene layer.
Fortunately, our results also show that it is not
necessary to adsorb several molecules to obtain 
large spin moments in graphene. 
A similar effect is obtained by adsorbing  
a {\it single} molecule that simultaneously establishes 
several C$-$C covalent bonds with the layer.
This simple observation might open a new route to engineer the
magnetism of graphene derivatives. In principle,
{\it molecules could be synthesized with the appropriate 
distances between anchoring groups} so, when adsorbed on graphene, 
all bonds will
be formed with atoms in the same
graphene sublattice.
The spin moment associated with the adsorption of such molecules
would be equal to the number of C$-$C bonds with 
graphene. Furthermore, the stability of this 
spin moment will be enhanced due to the strong FM interactions
between nearby localized moments in the same sublattice.

We have also analyzed our results for the magnetic interactions
within the framework of a classical Heisenberg 
model. The interactions are described by 
the model Hamiltonian 
$H= \sum_{i<j} J({\bf r}_{ij}){\bf S}_{i}{\bf S}_{j}$,
where $J({\bf r}_{ij})$ is the exchange constant 
and ${\bf S}_{i}$ is the local moment induced by 
covalent functionalization.
The expression for the angular  dependence of 
the exchange has been taken from the
RKKY-like coupling obtained analytically in Ref.~\cite{Saremi07}.
A simple $|r_{ij}|^\alpha$ behavior was assumed
for the distance dependece, and the exponent $\alpha$ was fitted
to our {\it ab initio} results in the case of AA-adsorption.
We find that the interaction between the chemisorbed molecules 
in graphene surface is long range and falls off slowly with the distance, 
roughly proportional to 
$J_{AA}({\bf r}_{ij})\sim |r_{ij}|^{-(1+\epsilon)}$ 
with $\epsilon \sim 0.20$.
[see the circles in Fig.~\ref{fig3}~(d) and (e)].
This distance decay 
agrees well with a recent theoretical
study using a $\pi$-tight-binding model~\cite{Levitov09}, 
where the interaction between adatoms on graphene was
shown to decay as 
the inverse of the distance.
In contrast, exchange interactions between  
substitutional Co impurities in graphene, 
a system that in principle can also be assimilated to a $\pi$-vacancy
or H adsorption,
were recently shown to decay 
much faster ($\sim |r_{ij}|^{-2.43}$)~\cite{Co-paper10}. 
This is probably related to the larger localization of the
spin moment in the latter case, associated with the 3$d$ Co contribution.

\section{Conclusions}

We have shown that 
covalent functionalization can be used 
to switch on the magnetism of graphene. Notwithstanding its structure and
chemical or biological activity, when a molecule
chemisorbs on graphene with the formation 
of a single C$-$C covalent bond between the adsorbate and the layer, 
a spin moment of 1.0~$\mu_{B}$ is always induced in the system.
This universal spin moment is almost exclusively 
localized in graphene,
and its 
origin can be traced back to the polarization of a defect state
that appears nearby the Fermi level.
A similar 
effect accompanies 
the functionalization of carbon nanotubes~\cite{Santos11}, and 
the chemisorption of hydrogen on graphene.
This behavior is also reminiscent of the polarization observed 
for a single vacancy in a
$\pi$-tight-binding description of the carbon monolayer.
Other types of bonding, mediated 
by species different from C, also show 
similar magnetic properties as far as the established bond 
is not strongly polar. If chemisorption is accompanied
by a substantial charge transfer between the adsorbate
and graphene, the spin polarization is reduced and eventually disappears, 
as it is the case of F adsorption.
However, even in this case, the low energy cost necessary
to polarize the system seems to indicate that the 
magnetic properties of fluorinated graphene can be tuned using 
moderate magnetic fields.~\cite{Hong11}

Similar to the moment formation, exchange interactions are also
found to be very weakly dependent on the nature of
the adsorbates. 
In particular, we have explicitly checked
that hydrogen atoms and methyl radicals give rise 
to the same magnetic couplings. 
Adsorbates in the same graphene sublattice (AA adsorption)
couple ferromagnetically with an exchange interaction
that decays almost inversely proportional 
to the distance $J_{AA}({\bf r}_{ij})\sim |r_{ij}|^{-(1+\epsilon)}$
with $\epsilon \sim 0.20$. In contrast, 
the larger effective electronic hoppings
between adsorbates at different
sublattices (AB adsorption) prevent magnetism from arising in that 
situation.
We have checked that the same results
are obtained irrespective of whether the C$-$C bonds are established
by the simultaneous adsorption of several molecules
or by the adsorption of a single larger molecule.
Therefore, 
the observed behavior is intrisic
to graphene and appears 
associated with the saturation of a given collection of p$_z$ 
orbitals in the layer. 
This might open new routes to engineer the
magnetism of graphene derivatives since, in principle, 
molecules can be produced with the appropriate
structural and chemical characteristics to simultaneously bond
to several atoms in the same graphene sublattice.
The spin moment arising from such an adsorption will be proportional
to the number of covalent bonds established and will have 
an enhanced stability due to the ferromagnetic interactions
between nearby saturations in the same sublattice.
Our findings are particularly attractive in the light of the recent
sucessful synthesys
of different graphene derivatives using surface chemical
routes~\cite{Mullen10,Mullen10_bis}.
Therefore, the synthesis of carbon nanostructures with functional groups
at predefined positions, starting from previously functionalized
monomers, seems plausible nowadays.

\ack
We acknowledge support from Basque
Departamento de Educaci\'on and the UPV/EHU (Grant
No. IT-366-07), the Spanish Ministerio de Educaci\'on
y Ciencia (Grant No. FIS2010-19609-CO2-02) 
and the ETORTEK program funded by the Basque Departamento de 
Industria and the Diputaci\'on Foral de Gipuzkoa.
DSP wants to acknowledge useful discussions with 
Dr. G. Teobaldi. 

\section*{References}

\providecommand{\newblock}{}


\begin{thebibliography}{10}
\bibitem{note1}
We can see in Fig.~\ref{fig1}(f)-(g) that other bands also suffer a splitting
  due to the spin polarization induced by the defect. However, those splittings
  are much smaller than that of the $p_z$ defect-state close to $E_{F}$
\end{thebibliography}
\end{document}